\documentclass[letterpaper,11pt]{article}

\usepackage[utf8]{inputenc}

\usepackage{times,helvet,courier}

\usepackage{amsmath,amssymb}

\usepackage{algorithm2e}

\usepackage{graphicx}
\usepackage{subfig}
\usepackage{multirow}
\usepackage{longtable}
\usepackage{array,booktabs}
\usepackage{float}

\usepackage{siunitx}

\usepackage{verbatim}

\title{Graph-Coarsening for Machine Learning Coarse-grained Molecular Dynamics}

\author{
  Soumya Mondal\textsuperscript{1,\textdagger}, 
  Subhanu Halder\textsuperscript{2,\textdagger}, 
  Debarchan Basu\textsuperscript{3},\\
  Sandeep Kumar\textsuperscript{2,3,*}, 
  Tarak Karmakar\textsuperscript{1,3,*}
}
\date{}  

\begin{document}
\maketitle

\begingroup
  \renewcommand\thefootnote{\textdagger}
  \footnotetext{These authors contributed equally to this work.}
  \renewcommand\thefootnote{*}
  \footnotetext{Corresponding authors: ksandeep@iitd.ac.in, tkarmakar@chemistry.iitd.ac.in}
\endgroup

\noindent
\textsuperscript{1}Department of Chemistry, Indian Institute of Technology Delhi, Hauz Khas, New Delhi 110016, India\\
\textsuperscript{2}Department of Electrical Engineering, Indian Institute of Technology Delhi, Hauz Khas, New Delhi 110016, India\\
\textsuperscript{3}Yardi School of AI, Indian Institute of Technology Delhi, Hauz Khas, New Delhi 110016, India


\begin{abstract}
Coarse‑grained (CG) molecular dynamics (MD) simulations can simulate large
molecular complexes over extended timescales by reducing degrees of freedom. A
critical step in CG modeling is the selection of the CG mapping algorithm,
which directly influences both accuracy and interpretability of the model.
Despite progress, the optimal strategy for coarse‑graining remains a
challenging task, highlighting the necessity for a comprehensive theoretical
framework. In this work, we present a graph‑based coarsening approach to
develop CG models. Coarse‑grained sites are obtained through edge contractions,
where nodes are merged based on a local variational cost metric while
preserving key spectral properties of the original graph. Furthermore, we
illustrate how Message Passing Atomic Cluster Expansion (MACE) can be applied
to generate ML‑CG potentials that are not only highly efficient but also
accurate. Our approach provides a bottom‑up, theoretically grounded
computational method for the development of systematically improvable CG
potentials.  \end{abstract}

\section{Introduction}
Atomistic molecular dynamics (MD) simulations are widely used to simulate
physical, chemical, and biological processes by integrating Newton's equation
of motion \cite{allen2017computer,frenkel2023understanding}. However, modern
advances in computational resources, atomistic MD simulations find limitations
in simulating many interesting phenomena that occur on longer timescales,
ranging from milliseconds to a few seconds, and involve a large number of
atoms, making them difficult to study.  A workaround is the use of large
integration timestep. However, this is often impractical due to the presence of
fast degrees of freedom associated with rugged free-energy landscapes, which
require fine temporal resolution to be accurately solved. Moreover, most
standard integrators are unable to maintain stability under large integration
timesteps. \cite{marrink2007martini,kmiecik2016coarse}. 
free-energy surface with small time steps becomes inefficient when a coarser
resolution is needed \cite{duschatko2024uncertainty}.

In addition, all-atom simulations are computationally demanding, as they
require the calculation of interatomic forces and the update of positions and
velocities for every degree of freedom at each time step. This high
computational cost further restricts the scale and duration of atomistic
simulations. 

To circumvent these limitations, multiscale modeling approaches such as
coarse-grained (CG) simulations are needed to extend simulation capabilities
for larger systems and longer timescales. The CG technique reduces the
dimensionality of the model system by mapping a set of atoms into CG beads
\cite{wu2023k,baratam2024sop,lesniewski2024insight,mondal2025unveiling,leonarski2013evolutionary,shmilovich2022temporally}.
Coarse-grained (CG) models improve computational efficiency and extend
accessible spatial and temporal scales, that can reliably reproduce structural
and thermodynamic properties of diverse systems, including molecular liquids,
polymers, proteins, and other bio-macromolecules
\cite{izvekov2005multiscale,craven2014structure,moore2014derivation,dunn2015bottom,moradzadeh2019transfer,narros2010influence,craven2014effective,zhang2013solvent,ricci2023integrating,shih2006coarse,saunders2013coarse,noid2013perspective}.
The choice of the bead mapping protocol determines the fate of the CG
simulations \cite{kohler2023flow}. Currently, there is no generic universal
theory for defining CG beads for any given molecular system
\cite{foley2020exploring,boninsegna2018data}. Despite significant efforts,
developing accurate and efficient CG models still remains a challenge
\cite{peng2023openmscg}. Generally, the criteria for selecting a CG mapping
algorithm are based on a priori considerations and chemical intuitions
\cite{li2020graph,wang2019coarse}. For example, the widely used MARTINI CG
model uses a four-to-one mapping protocol \cite{souza2021martini}. Another
important CG mapping scheme for proteins and peptides is to select only the
$C_{\alpha}$ atom for each amino acid \cite{majewski2023machine}. However, the
selections of mapping schemes are not based on any thermodynamic or theoretical
arguments. In recent years, significant effort has been devoted to developing a
systematic and automated CG mapping scheme for a molecule
\cite{stroh2023cgcompiler,potter2021automated}. Automation of the bead mapping
scheme is essential to enhance the scalability and transferability of the CG
model.  Wang \textit{et al.} developed a generative AI framework based on
auto-encoders (AEs) to learn optimal coarse-grained variables or parameters
\cite{wang2019coarse}. The same group proposed Coarse-Grained Variational
Auto-Encoder (CGVAE) to generate coarsen representation from fine-grain
coordinates \cite{wang2022generative}. Reidenbach and coworkers proposed the
CoarsenConf architecture, which coarsened any given molecule based on an
SE(3)-equivariant hierarchical variational autoencoder
\cite{reidenbach2024coarsenconf}. Giulini \textit{et al.} proposed an
entropy-based mapping scheme to simplify the CG mapping scheme for biomolecules
\cite{giulini2020information}. Their CG model focused on preserving most of the
information in a low-resolution CG representation.


The application of graph theory to chemical systems has experienced remarkable
growth over the past decades, driven by the inherent ability of graphs to
naturally represent molecular structures where atoms serve as vertices and
chemical bonds as edges \cite{iyengar2024reformulation,pietrucci2011graph}. The
success of graph-based molecular representations has naturally extended to
addressing one of the most fundamental challenges in molecular simulation: the
development of efficient coarse-grained representations. The integration of
molecular graphs with automated bead definition represents a significant
advancement in coarse-grained molecular dynamics methodology. Graph-based
coarse-graining approaches treat molecular systems as mathematical graphs and
systematically derive coarse-grained representations through algorithmic graph
operations such as edge contractions and node
\cite{zhang2018deepcg,bejagam2018machine,lemke2017neural,wang2019machine,boninsegna2015investigating,thaler2022deep}.
This methodology offers several compelling advantages: it provides unambiguous
and reproducible mapping schemes, preserves chemical topology automatically,
and can generate hierarchical representations with varying levels of
resolution.  Recently, Webb \textit{et al.} applied a spectral and progressive
grouping scheme on molecular graphs to generate the CG representation of a
given molecule \cite{webb2018graph}. Chakraborty and coworkers developed a CG
mapping method based on hierarchical graphs \cite{chakraborty2018encoding}.
Recently the same group proposed a GNN-based CG mapping predictor named Deep
Supervised Graph Partitioning Model (DSGPM) \cite{li2020graph}.

Once a CG mapping function is defined, the model energy function can be
constructed using bottom-up approaches, which reproduce fine-grained
statistics, or top-down strategies, which fit experimental observables
\cite{jin2022bottom,navarro2023top}. A coarse-grained model is governed by the
many-body potential of mean force (PMF), which encapsulates the system’s
configurational free energy in a reduced phase space. However, accurately
constructing a many-body PMF that preserves the structure, thermodynamics, and
kinetics of the coarse-grained system remains challenging. Due to the
complexity of the PMF, machine learning (ML) force fields have gained
importance in efficiently predicting the accurate potential energy functions
for use in ML potential-based classical MD simulations by training on a large
dataset
\cite{chen2021machine,unke2019physnet,unke2022accurate,unke2019physnet,noe2020machine}.
Extensive research has been conducted in the field of ML potentials. A key
advantage of ML potentials is their ability to capture complex many-body atomic
interactions \cite{wang2021multi}. The success of all-atom ML force fields as
surrogates for ab initio dynamics has sparked interest in using similar methods
for PMF modeling in CG systems. These methods use a force-matching technique
that reproduces all-atom PMF in the limit of sufficient sampling space
\cite{noid2008multiscale}. Numerous ML-CG models have emerged in recent years,
but most depend on predefined energy terms to ensure stability and accuracy
\cite{wang2019machine,husic2020coarse,chen2021machine,durumeric2023machine,charron2023navigating}.
Recently, Wang \textit{et al.} applied the ACE method to construct
computationally efficient many-body coarse-grained potential
\cite{wang2025many}.
In particular, the MACE \cite{batatia2022mace,poltavsky10crash} method provides
systematic, flexible yet computationally efficient solutions for this purpose.
Even though the MACE method has been widely used to generate interatomic
potentials, it has not been applied to the ML-CG context to the best of our
knowledge.

In this work, we coarsen the molecular graph been merging some of the nodes of
the graph and creating a supernode based on the local variation algorithm,
which has not been designed heuristically, but optimizes the objective of
spectral similarity. To construct a coarsened molecular graph for ML-CG, we
define neighborhood and clique-based candidate sets that greedily choose to
contract based on minimum local variation. Subsequently, we utilize the
force-matching scheme to optimize the MACE model parameters for the CG
molecular graph. In this scheme, one can directly derive free energy gradients
from standard atomic forces, often called “instantaneous forces” in the CG MD
literature, sampled from MD simulation trajectories \cite{husic2020coarse}. We
tested the MACE-CG method on three prototypical model systems: Aspirin,
Azobenzene, and 3BPA with increasing conformational flexibility. In addition,
we demonstrate the MACE-CG model's capability to accurately represent
equilibrium properties such as bond lengths and radial distribution functions
(RDFs). We demonstrate the efficacy and validity of local variation based
graph-coarsening method by analyzing the equilibrium properties and then
comparing them with the respective ground truth or reference data. Our results
demonstrate the capabilities of the advanced graph coarsening algorithm in the
ML-CG context.

\section{Methods}
\section*{Methods}
The machine learning-based force fields received significant attention in the
past few years to learn potential energy surfaces (PESs). Several machine
learning architectures have been developed to address problems in ML-CG
simulations. Popular approaches such as relative entropy minimization and
variational force matching are usually used to train the CG models. The latter
one is the widely adopted approach in the bottom-up CG perspective. The force
matching scheme reduces the discrepancy between the predicted CG forces and the
true atomistic forces mapped onto the CG space. In the following section, we
summarize the theoretical foundations of force-matching approach.

\subsection*{Force Matching}
Consider a potential energy function $U(\mathbf{r}^{(i)};\Theta)$ for the
system containing  $N$ number of atoms with atom coordinates $\mathbf{r}^{(i)}$
and tunable parameters $\Theta$. Neural Network Potentials have emerged as
powerful tools to parameterize this energy function by computing the gradient
of the learned free energy function $\nabla_{\mathbf{r}^{(i)}}
U(\mathbf{r}^{(i)}; \Theta)$.  The NN parameters $\Theta$ are optimized to
minimize the square error between predicted and the true forces via the loss
function.
\begin{equation}
\mathcal{L}(\Theta) = \frac{1}{3MN} \sum_{i=1}^{M} 
\left\| 
\underbrace{\mathbf{F}^{(i)}}_{\text{True forces}} 
+ 
\underbrace{\nabla_{\mathbf{r}^{(i)}} U(\mathbf{r}^{(i)}; \Theta)}_{\text{(Negative) predicted forces}} 
\right\|_2^2
\label{Force_match_original}
\end{equation}
Here, \(M\) is the number of frames in thedataset, \(N\) is the number of atoms
in the system, and \(\mathbf{F}^{(i)}\) denotes the reference (true) forces.

Force-matching approach to parameterize CG potential was proposed in Ref
(\cite{izvekov2005multiscale}) to reproduce the structural correlation present
in the ground-truth dataset.  In CG setup, we need to define a mapping matrix
or coarsening operator $P: \mathbb{R}^{N} \to \mathbb{R}^{n},\; \mathbf{r}
\rightarrow \mathbf{R}$
 to project fine grained state R onto a lower dimesnional coarsened space
\(\mathbf{r}_{r}^{(i)}= Pr^{(i)} \in \mathbb{R}^{n}\) in reduced
configurational space from high-dimensional atomistic representation and
\textit{instantaneous } forces projected on the CG coordinates (also called
\textit{local mean force}) \(\mathbf{F}_{r}^{(i)}= P_F \mathbf{F}^{(i)} \),
$P_F$ projects the atomic forces into the the CG space. The force projection
operator is typically defined as: \begin{equation}
    P_F = (PP^T)^{-1}P
    \label{Projectio P}
 \end{equation}
which provides a least-squares optimal mapping of atomistic forces to the CG
subspace~\cite{ciccotti2008projection}. However, this formulation can be
computationally demanding and sensitive to the structure of $P$. To address
this, we adopt a binary assignment matrix, where each atom is uniquely mapped
to a CG site. Under this discrete partitioning, $PP^T= I$  becomes the
identity, simplifying the projection Eq.\ref{Projectio P} to  $P=P_F$. This not
only enhances computational efficiency but also ensures physical
interpretability and preserves key symmetry properties—such as translational
and rotational equivariance—that are essential for training CG potentials using
equivariant neural networks like MACE \cite{batatia2022mace,
kovacs2023evaluation}. Importantly, this formulation integrates naturally with
graph-based approaches, where atoms and CG sites are represented as nodes in
fine and coarse graphs, respectively. The operator $P$ thus defines a graph
coarsening map that preserves local structure and neighborhood information,
both of which are critical for learning accurate many-body interactions in
molecular systems.

CG model is often defined by a CG energy function \( U(P\mathbf{r}^{(i)};
\Theta) \). Assume we constructed a dataset of \( M \) coarse-grained
configurations by applying the mapping function $P$ to each atomistic
configuration \( \mathbf{r}^{(i)} \in \mathbb{R}^N \) present in the dataset.
To compute the forces on the coarse-grained beads, we take the negative
gradient of \( U \) with respect to the coarse-grained coordinates \(
\mathbf{r}^{(i)} \in \mathbb{R}^n\), i.e. for each configuration $i$, $-\nabla
U(P\mathbf{r}^{(i)}; \Theta)= -\nabla_{\mathbf{r}^{(i)}} U(\mathbf{r}^{(i)};
\Theta) \in \mathbb{R}^n$

In MACE, a molecule is treated as a molecular graph, where atoms or beads are
represented by nodes. Through equivariant message-passing layers, MACE builds a
many-body description of the molecule using the Atomic Cluster Expansion (ACE)
framework~\cite{niblett2025transferability} by combining radial and angular
information as messages. The energy associated with each CG site, \(
U_{\mathrm{MACE}}(\{\mathbf{r}_{r}\}) \), is predicted using a multilayer
perceptron~\cite{gel}. The CG force field is then trained by minimizing the
instantaneous force matching loss: \begin{equation}\label{eq:force_loss}
\mathcal{L}(\Theta)
\;=\;
\frac{1}{3 M n} \sum_{i=1}^{M} 
\Bigl\|\,
\underbrace{P \,\mathbf{F}^{(i)}}_{\text{CG mapped atomistic forces}}
\;+\;
\underbrace{\nabla_{\mathbf{r}}\,U_{\mathrm{MACE}}(P\,\mathbf{r}^{(i)}; \Theta)}_{\text{(Negative) forces predicted from CG model})}
\Bigr\|_{2}^{2},
\end{equation}

Minimizing \(\mathcal{L}(\Theta)\) trains the MACE network so that
\[
-\nabla_{\mathbf{r}}\,U_{\mathrm{MACE}}\bigl(P\,\mathbf{r}; \Theta\bigr)
\;\approx\; P\,\mathbf{F}
\]
A detailed description of MACE architecture can be found in
Ref(\cite{kovacs2023evaluation}). The key MACE parameters for the MACE training
are reported in the Results section.  In the following section, we summarize
the hierarchical framework for defining the coarsening matrix $P$.

\subsection*{Molecular Graph Representation} We define a molecular graph $G =
(V_0, E_0, W_0)$, where $V_0$ is a set of vertices corresponding to the heavy
atoms $i = 1, \ldots, N$ of a molecule in its atomistic representation, $E_0$
is a set of edges $e_{ij}$ connecting pairs of atoms $i$ and $j$ within a
cutoff distance, and $W_0$ is a $N \times N$ weight matrix whose elements are
$w_{ij}$. From the atomistic graph, we successively derive coarser graph
representations, $G_\ell = (V_\ell, E_\ell, W_\ell)$, l=1,2,3,...c by
contracting the vertices in $V_\ell$

The $i^{\text{th}}$ vertex of the molecular graph corresponding to the
$i^{\text{th}}$ heavy atom in the molecule is represented with its coordinates
$\vec{\mathbf{r}}_i = (x_i, y_i, z_i)$ and force $\vec{\mathbf{F}}_i = (F_{xi},
F_{yi}, F_{zi})$. An edge $e_{ij} \in E$ is inserted whenever the Euclidean
distance $d_{ij} = \|\vec{\mathbf{r}}_i - \vec{\mathbf{r}}_j\|$ is less than a
cutoff distance $d_{\text{cut}}$, which is chosen to include all covalent bonds
and short non-bonded contacts. The weight $w_{ij} \in W$ on each edge is
defined as: \[
w_{ij} = \exp\left[-\alpha (d_{ij} - d_0)\right]
\]
where $d_0$ is the reference bond length, and $\alpha$ is a damping parameter,
whose value is chosen such that bonded atoms and close contacts have $w_{ij}
\approx 1$ while the rest have weights zero.

Now we can write the combinational (graph) Laplacian $L \in \mathbb{R}^{N
\times N}$ as: \[
L = D - W_0
\]
where $D$ is the degree matrix, $D = [d_{ij}]$, whose elements $d_{ii}$ are
defined as the sum of diagonal elements of the weight matrix $W_0$ as: $d_{ii}
= \sum_{j: e_{ij} \in E} w_{ij}$

\begin{figure}[!ht]
    \centering
    \includegraphics[width=0.7\linewidth]{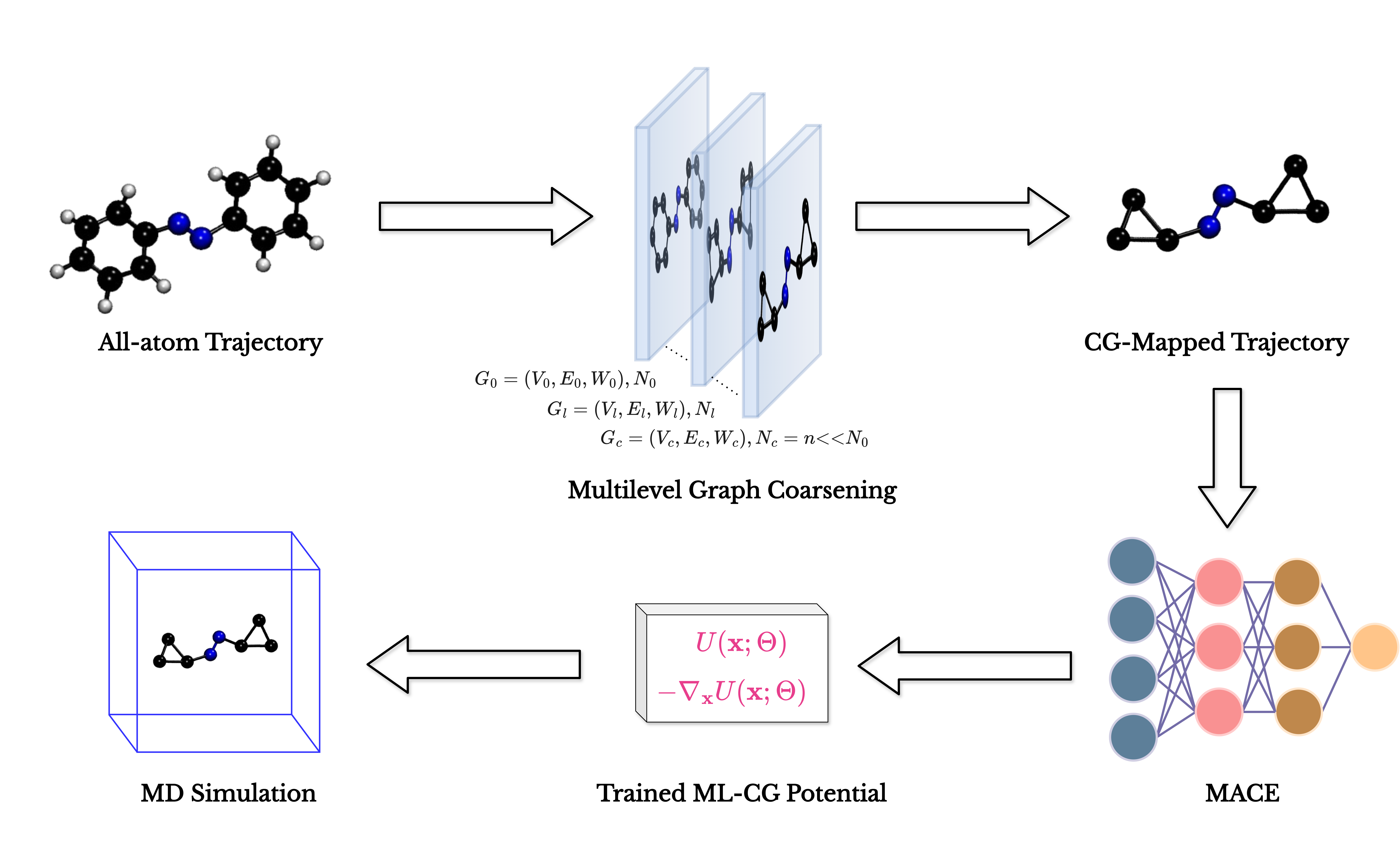}
    \caption{The overall framework of multilevel graph-coarsening for bottom-up
ML-CG molecular dynamics simulations. A multilevel graph coarsening technique
is applied to the no-hydrogen atoms (noh) dataset to reduce it to a
low-dimensional CG representation. Then MACE architecture is trained using the
coordinates and forces from the CG-mapped trajectory to predict the ML-CG
potential.} \label{fig:multi}
\end{figure}

\subsection*{Coarse‐Graining Mapping Procedure}

A good force matching performance in coarse-grained (CG) models depends heavily
on the choice of the mapping matrix \(P\), which maps atoms to CG beads. There
are several ways to choose this matrix \cite{10.5555/3618408.3619148,
JMLR:v24:22-1085, pmlr-v244-kumar24a,
10.1145/3701716.3715522,pmlr-v108-jin20a}. Some methods are based on
optimization, while others rely on deep learning. However, deep learning
methods are not deterministic, meaning they can produce different results on
different runs. They also lack interpretability and do not offer clear control
over how atoms are merged into beads. In addition, these methods require
training and GPU resources, making them expensive and less transparent.

To define the coarsening matrix in our work, we use the framework of
\textit{Multilevel Graph Reduction with Spectral and Cut Guarantees}
\cite{loukas2019graph}, adapting it to the context of CG molecular dynamics.

The main idea is to represent a molecule as a graph \(G_0 = (V_0, E_0, W_0)\),
where each node in \(V_0\) corresponds to a heavy atom, and edges in \(E_0\)
connect atoms that are within a cutoff distance. We then apply a multilevel
graph coarsening procedure to derive successively reduced graphs \(G_{\ell} =
(V_{\ell}, E_{\ell}, W_{\ell})\). At each level, atoms are grouped into
connected clusters called \emph{contraction sets}, using the Local Variation
Neighborhood (LVN) and Local Variation Cliques(LVC) criterion. This coarsening
process is repeated until the number of coarse-grained nodes reaches a desired
target \(n\), which is controlled by a \textit{coarsening ratio}: \[
r = 1 - \frac{n}{N}
\]
where \(N\) is the number of nodes (vertices) in the original graph, and \(n\)
is the number of nodes in the coarsened graph. An overview of the full
coarsening pipeline is shown in Figure~\ref{fig:multi}.

To learn a chemically meaningful coarsened graph and a well-structured
coarsening matrix \(P\), we adopt a multilevel graph reduction approach that
ensures both structural fidelity and spectral consistency across coarsening
levels.  At each level \(\ell = 1,2,\ldots,c\), we coarsen \(G_{\ell-1} =
(V_{\ell-1}, E_{\ell-1}, W_{\ell-1})\) to a reduced graph \(G_{\ell} =
(V_{\ell}, E_{\ell}, W_{\ell})\) with \(|V_{\ell}| = N_{\ell} < N_{\ell-1}\).
During coarsening, one needs to ensure that the graph's spectral properties are
preserved. To minimize the restricted (refers to a subspace vector) spectral
approximation error, we maintain a subspace basis
\(B_{\ell-1}\in\mathbb{R}^{N_{\ell-1}\times k}\), typically chosen as the first
\(k\) eigenvectors of the Laplacian (\(L_{\ell-1}\)).  
  
We begin by constructing a candidate family \(\mathcal{F}_\ell = \{C_1, C_2,
C_3, \dots\}\) of connected vertex subsets \(C_{i}\) in \(V_{\ell-1}\), where
\(C_{i}\)s are called "candidate sets" or "contraction sets". Each candidate
\(C \in \mathcal{F}_\ell\) has size \(N_C = |C| \ge 2\), i.e., it contains at
least two vertices. Each contraction set \(C \in \mathcal{F}_\ell\) is a
candidate to be merged into a single supernode at level \(\ell\), thereby
forming the coarser vertex set of \(v' \in V_\ell\) from \(v \in V_{\ell-1}\).
Selecting a subset \(\mathcal{P}_\ell \subseteq \mathcal{F}_\ell\) defines the
coarsening at level \(\ell\). 

\begin{figure}[!h]
    \centering
    \includegraphics[width=1.0\linewidth]{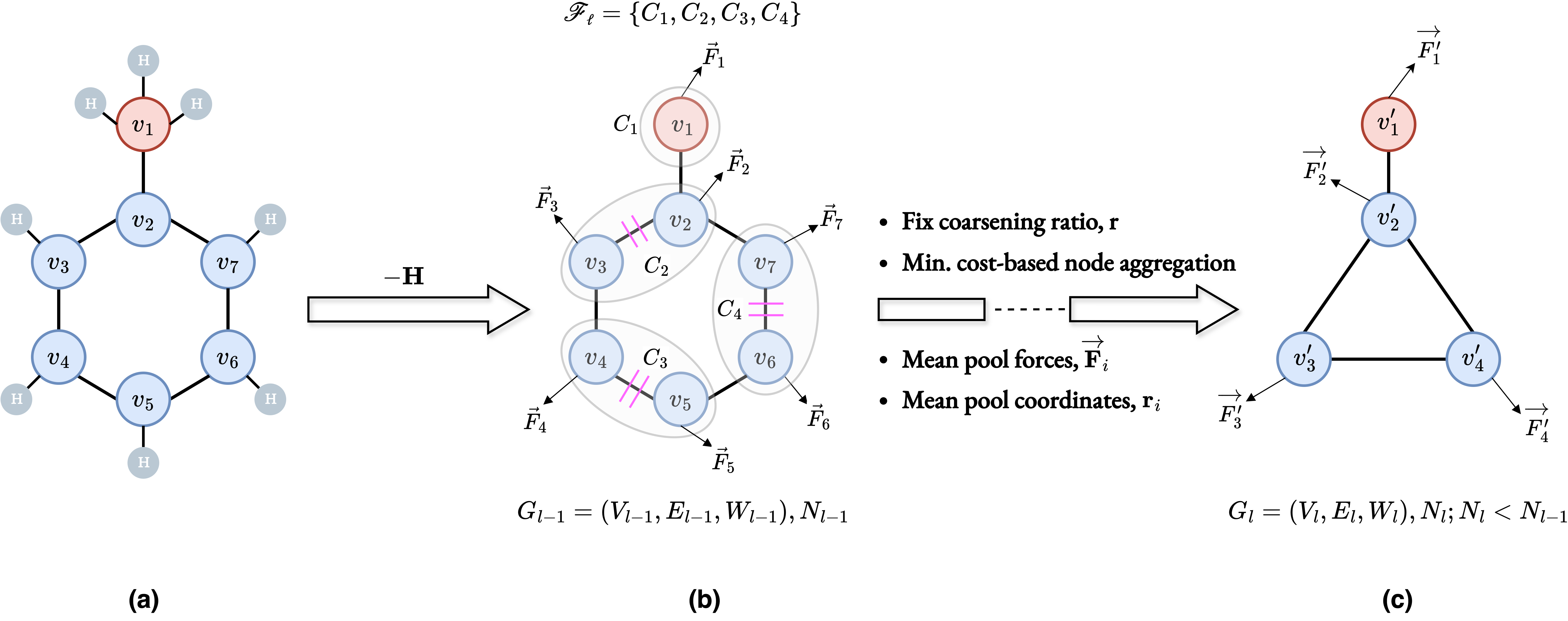}
    \caption{Toy model for developing a coarsened graph using the multilevel
graph coarsening technique. a) All-atom representation b) Nonhydrogenic
representation. The grey discs denote contraction elements from the candidate
sets present in the candidate family ($\mathcal{F}_{\ell}$) based on local
variational cost. $F_{i}$ denotes the force on each node. c). Coarsened graph
with target nodes and reduced edges} \label{fig:coarsen}
\end{figure}

We employed the following two strategies to construct candidate contraction
families \(\mathcal{F}_\ell\) (See fig \ref{fig:coarsen}): \begin{itemize}
    \item Local Variation Neighborhood (LVN): A common effective strategy is to
form candidate sets by taking the one-hop neighborhood of each vertex,
including the vertex itself.  \[ \mathcal{F}_\ell^{\text{LVN}} = \left\{ \{v\}
\cup \mathcal{N}(v) : v \in V_{\ell-1} \right\}, \quad \text{where }
\mathcal{N}(v) = \left\{ u \in V_{\ell-1} : (u,v) \in E_{\ell-1} \right\}.
\]
    \item Local Variation Cliques (LVC):
Following the findings of Ref(~\cite{ghosh2008variations}), graph cliques are
particularly relevant in the study of allosteric regulation, where residue
groups often act in concert to mediate long-range communication across a
protein. Therefore, we define: \[
\mathcal{F}_\ell^{\text{LVC}} = \left\{ C \subseteq V_{\ell-1} : C \in \mathrm{Cliques}(G_{\ell-1}),\ |C| \ge 2 \right\},
\]
where each \(C\) is a maximal clique identified from the underlying graph
structure \(G_{\ell-1}\), and \(\mathrm{Cliques} (G)\) denotes the set of all
such cliques.

\end{itemize}

To evaluate the quality of a coarsening step \( G_{l-1} \rightarrow G_{l} \),
we use the variation cost \(\sigma_\ell\), defined as \[
\sigma_\ell  = \left\| S_{\ell-1} \Pi^\perp_\ell A_{\ell-1} \right\|_2,
\]
where \(L_{\ell-1} = S_{\ell-1}^\top S_{\ell-1}\) defines an inner product
where, the matrix \( S_{\ell-1} \in \mathbb{R}^{n_{\ell-1} \times n_{\ell}} \)
maps the vectors from the coarse level \(\ell\) back to the fine level \(\ell -
1\), and \(\Pi^\perp_\ell = I - P_\ell^\top P_\ell\) is the orthogonal
projection onto the complement of the subspace \(\mathbb{R}\) (i.e., the span
of the leading eigenvectors) we aim to preserve and $P$ be a coarsening matrix.
Minimizing \(\sigma_\ell\) ensures that the coarse operator \(A_{\ell-1}\)
introduces minimal distortion in directions orthogonal to \(\mathbb{R}\),
thereby preserving the spectral structure of the original graph. This framework
enables a principled, multi-level reduction scheme that selects contraction
sets with low local variation cost to approximate the global structure
effectively.

For each \(C\), we define the local variation cost,
\begin{equation}\label{eq:cost}
\mathrm{cost}_{\ell}(C) \;=\; 
\frac{\bigl\|\,S_{\ell-1}\,\Pi^{\perp}_{C}\,A_{\ell-1}\bigr\|_{2}^{2}}{|C| - 1},
\end{equation}

 Intuitively, \(\|\,S\,\Pi^{\perp}_{C}\,A\|_{2}\) measures the worst‐case
spectral distortion induced by contracting \(C\). The matrix \( A_{\ell-1} \)
captures the structure of the target Laplacian with respect to the subspace
\(\mathbb{R}\). In the beginning \( A_{l-1} = A_{0} \), it is defined by
\[
A_0 = \mathcal{V} \mathcal{V}^\top \sqrt{L},
\]
where \(\mathcal{V} \in \mathbb{R}^{N \times n}\) is an orthonormal basis for
\(\mathbb{R}\). For levels \(\ell > 1\), it is recursively defined as \[
A_{\ell-1} = B_{\ell-1} \left( B_{\ell-1}^\top L_{\ell-1} B_{\ell-1} \right)^{+1/2},
\quad \text{with} \quad B_{\ell-1} = P_{\ell-1} B_{\ell-2}, \quad B_0 = A_0.
\]

Based on this intuition, the algorithm proceeds greedily: At each step, it
contracts the lowest‐cost candidates \(C\) whose vertices have not yet been
assigned, provided the resulting per‐level variation \(\sigma_{\ell}\) does not
exceed the prescribed threshold \(\sigma_{\max}  \leq \prod_\ell (1 +
\sigma_\ell) - 1\). If a candidate \(C\) overlaps previously selected vertices,
it is pruned to \(C' = C\setminus \{\text{marked}\}\), its cost
\(\mathrm{cost}_{\ell}(C)\) is recomputed, and it is reinserted if \(|C'|\ge
2\).  Singleton sets are used for any vertices that remain unassigned after the
loop.  

Once \(\mathcal{P}_\ell\) is determined, one obtains the coarsened Laplacian
\[
L_{\ell} \;=\; \mathcal{P}_{\ell}^{T} \, L_{\ell-1} \, \mathcal{P}_{\ell},
\]
which remains a combinatorial Laplacian whose edge weights satisfy
\[
w_{pq}^{(\ell)} \;=\; \sum_{\,i\in C_{p}} \sum_{\,j\in C_{q}} w_{ij}^{(\ell-1)}, 
\]
ensuring that \(L_{\ell}\) preserves the cut‐weights exactly \cite{loukas2019graph}.  

After \(c\) levels of coarsening, we obtain graph \(G_{c} =
(V_{c},E_{c},W_{c})\) with \(\lvert V_{c}\rvert = N_{c} = n\).  The overall
coarsening matrix is
\begin{equation}
P \;=\; P_{c} \,P_{c-1} \cdots P_{1} \;\in\;\mathbb{R}^{n\times N}.
\label{projection def}
\end{equation}

Each row \(r\) of \(P\) corresponds to a bead in the CG model, identified by
the final contraction set \(C_{r}\), and \(\sum_{i=1}^{N} P_{r,i} = 1\).
Consequently, the coarse‐grained coordinate of bead \(r\) is the centroid
(center-of-geometry) of its constituent atoms: \begin{equation}
    \mathbf{r}_{r}
\;=\;
\frac{1}{|C_{r}|}
\sum_{i \in C_{r}} \mathbb{R}_{i},
\quad r = 1,\dots,n.
\label{cCordinate}
\end{equation}
Similarly, the coarse‐grained force on the bead \(r\) is
\begin{equation}
\mathbf{F}_{r} \;=\; \frac{1}{|C_{r}|}\sum_{i\in C_{r}} \mathbf{F}_{i}.
\label{CForce}
\end{equation}

\section{Results}

In the following sub-sections, the performance of the ML CG potential for
different molecules are explored. All simulations were performed using the
Atomic Simulation Environment (ASE) package \cite{larsen2017atomic}. The
simulations were carried out in the canonical (NVT) ensemble at the
temperatures in accordance with the datasets (500 K for Aspirin and Azobenzene
and 300 K for 3-(benzyloxy)pyridin-2-amine (3BPA)). 

\subsection*{Aspirin}
\begin{figure}[!h]
    \centering
    \includegraphics[width=0.6\linewidth]{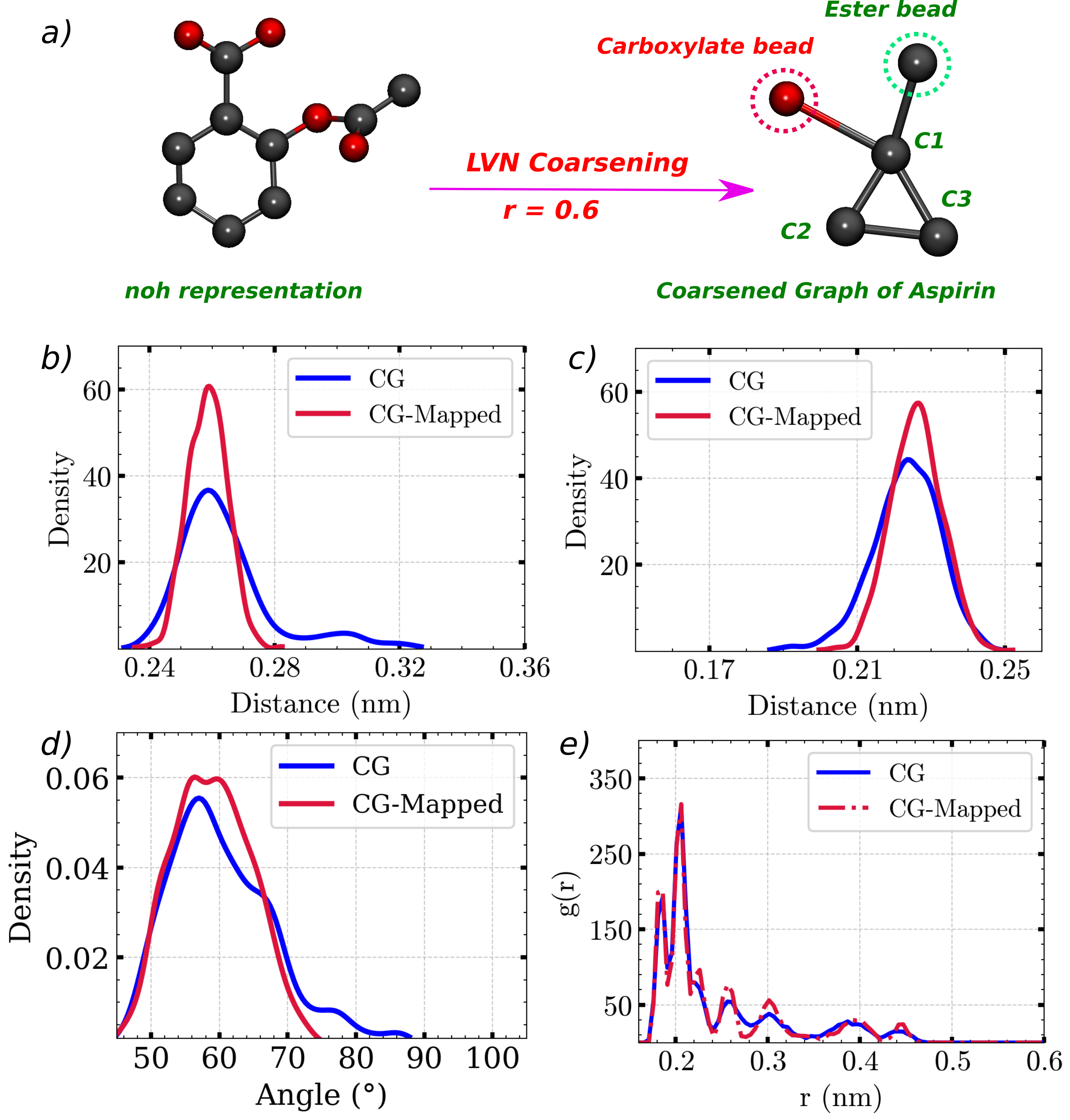}
    \caption{a). Demonstration of the coarsening process for the aspirin
molecule using LVN coarsening algorithm. Probability distribution of b).
distance between C1 bead and carboxylate bead c). distance between C1 and ester
bead, and d). density distribution of angle ($\theta$) between two vectors
passing through carboxylic acid and ester groups. e). g(r) of atomic density.
Blue line for CG trajectory and  red line for CG-Mapped trajectory}
\label{fig:asp}
\end{figure}

First, we have selected the aspirin drug  a common pharmacutical, as a model
system in order to validate the proposed unsupervised  multilevel graph
coarsening technique. In our study the training and testing files for aspirin
were taken from the MD17 data set (a popular, widely used dataset)
\cite{christensen2020role} to cover a wide range of configurations of aspirin.
The MD17 benchmark data set consists of configurations of 10 small organic
molecules obtained from vacuum-sampled DFT-MD simulations at 500 K. For each
system (including all-atom (AA),  noh, and CG), the MACE architecture is
trained until the instantaneous force-matching loss is minimized, as mentioned
in the Methods section. Further details of hyperparameters for aspirin at each
resolution are provided in the SI. In AA and noh representations, the MACE is
trained for 500 and 300 epochs, respectively. While for CG representation, we
performed the training upto 100 epochs. Further details of training parameters
are provided in Table I in the SI.  Here, we carried out three sets of
independent simulations. In the first set, an all-atom simulation was carried
out using the MACE potential by placing a single aspirin in a cubic box of 2.4
nm. In the second and third sets, the noh and CG representations of aspirin
were simulated at 500 K using MACE-based ML potential. In each system, the
simulations were carried out with a timestep of 1 fs at 500 K.  We train the
MACE architecture on 950 configurations and validate on 50 configurations for
each representation. The noh representation of the aspirin molecule is
coarsened into a five-particle system using the LVN algorithm by fixing the
coarsening ratio at 0.6, as shown in Figure \ref{fig:asp}a. 

Interestingly, the neural network nicely captures the bond length distribution
for the coarsened model and reproduces the pair distribution function
accurately in each simulation. To highlight the benefits of the multilevel
coarsening algorithm, we compare the probability distribution of different
bonds between the CG trajectory and the CG-mapped trajectory (ground-truth).
The probability distribution of the distance between bead C1 and the
carboxylate bead exhibits a dominant peak at around 0.26 nm in both CG and
CG-Mapped trajectory as shown in Figure \ref{fig:asp}b. While Figure
\ref{fig:asp}c shows that the average distance between bead C1 and the ester
group appears at 0.22 nm in both the CG trajectory and the ground truth
dataset. Similarly, in AA and noh systems, we measured the distance between the
center-of-mass (COM) of the carboxylate/ester group from the selected atom, as
shown in Figure 1 in the SI. Figure 2a in the SI shows that the MACE
architecture nicely captures the C1 atom-carboxylic acid group distance in AA
simulation. While C2 atom-ester group bond distance in AA simulation appears at
0.19 nm as shown in Figure 1b in the SI. We also investigate the C1
atom-carboxylic acid and C2 atom-ester bond length distribution in noh system.
Figure 3a,b in the SI shows nice agreement between the underlying noh
trajectory and the ground truth dataset. To quantify the orientation of the
carboxylate and ester group in aspirin, we define an angle $(\theta)$ by
selecting two vectors passing through the carboxylate and ester bead as shown
in Figure 1 in the SI. Apparently, Figure \ref{fig:asp}d shows the probability
distribution of $\theta$, where the dominant peak appears at $60^\circ$ in both
cases, suggesting the capability of the ML-CG model to reproduce the structural
correlation present in the training dataset. Moreover, Figure 2c,d and 3c,d in
the SI depict the schematic representation of $\theta$ angle and its density
distribution in both AA and noh systems. The probability density profile of
$\theta$ obtained via AA and noh simulation nicely reproduce the density
distribution of their respective ground truth dataset. Radial distribution
function (g(r)) is often used to characterize the degree of local ordering in a
system \cite{shinkle2024thermodynamic}. In Figure \ref{fig:asp}e, we compared
the g(r) of atomic density between the CG trajectory and the CG-Mapped
trajectory (ground truth) at 500 K. The g(r) profile obtained from our CG
simulation reflects the ability of the MACE CG potential to capture structural
correlations similar to the ground truth dataset. Interestingly, Figure 2e and
3e in the SI depict the g(r) of AA and noh simulation compared to those of the
mapped reference or ground truth dataset, suggesting both the methods perform
well. Moreover, Jensen–Shannon Divergence (JSD) of the pair distribution
function (g(r)) is performed to quantify the similarity between the CG
trajectory and the mapped CG trajectory (ground-truth) as shows in Table III in
the SI. Table III in the SI shows MACE architecture performs better on each set
of simulations with a very low JSD value. Each ML-CG potential does a good job
of capturing the structural correlations of aspirin and allows us to evaluate
the impact of different levels of coarsening on reproducing all-atom system
properties.  

\subsection*{Azobenzene}
\begin{figure}[!ht]
    \centering
    \includegraphics[width=0.6\linewidth]{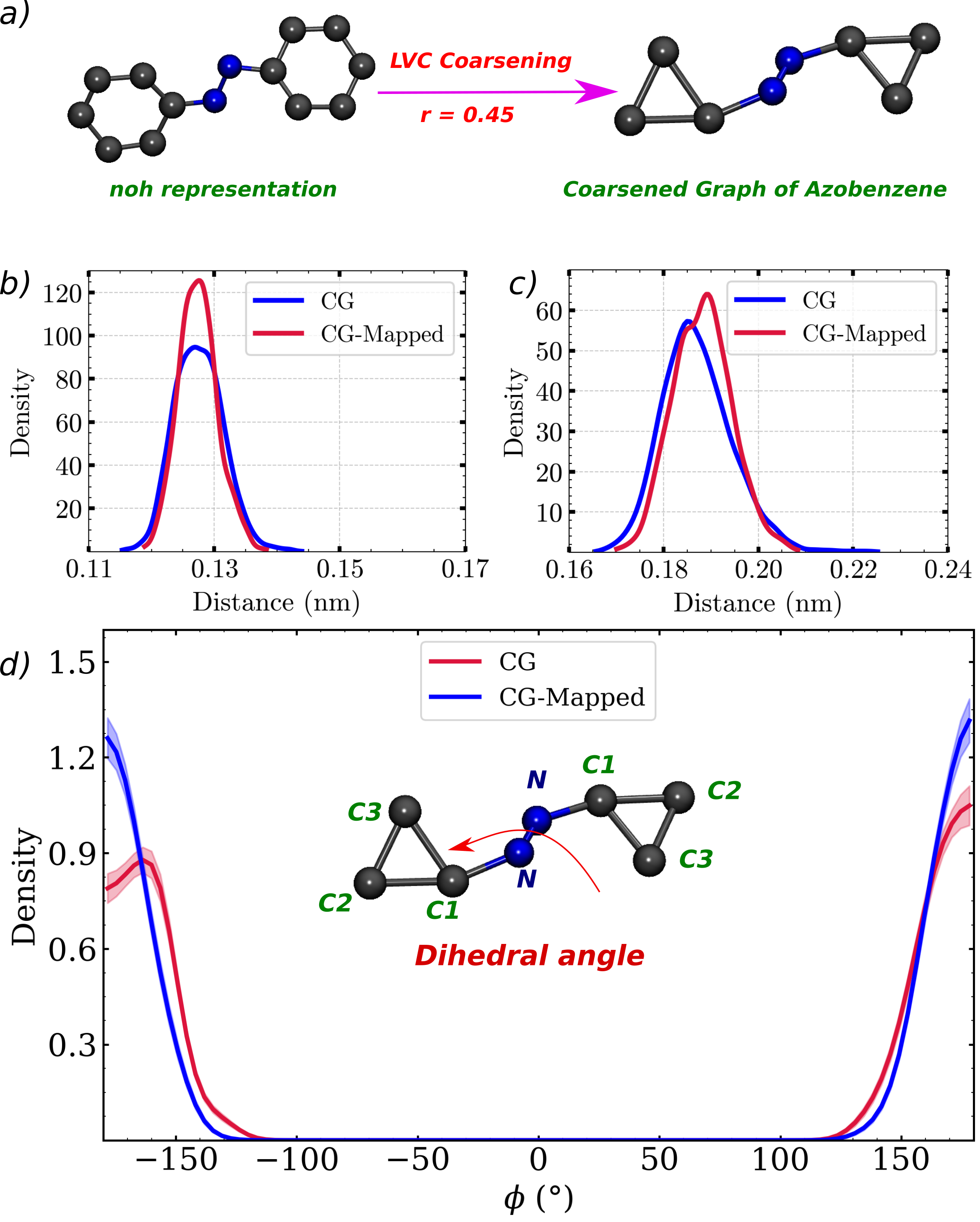}
    \caption{a). Demonstration of the coarsening strategy for the aspirin
molecule using the LVN coarsening algorithm. Probability distribution of b).
N=N bond distance c). distance between C1 and N atoms, and d). Dihedral angle
density distribution. Blue line for CG trajectory and red line for CG-Mapped
data} \label{fig:azo}
\end{figure}

To establish a baseline of the multilevel-coarsening algorithm, we selected
azobenzene, a large, flexible organic molecule, from the MD17 benchmark
dataset. The C–N–N–C rotatable dihedral angle shows a complex dihedral
potential energy surface with several local minima. Here, we train the MACE
architecture using 950 configurations sampled at 500 K and validate on 50
configurations in each case using the instantaneous force matching scheme
mentioned in the method section. For AA and noh representation, we train the
MACE architecture upto 400 and 200, respectively. For CG representation, the
MACE was train upto 160 epochs. Further details of hyperparameters are provided
in Table II in the SI. Specifically, we employ the unsupervised local
variational cliques (LVC) algorithm to construct the CG representation for
azobenzene by fixing the coarsening ratio at 0.45, as shown in Figure
\ref{fig:azo}a using coordinates and forces from the noh representation of
azobenzene.  
In the coarsened representation of azobenzene, the aromatic phenyl moieties are
represented by three CG beads. Interestingly, the LVC algorithm preserves the
ring-like geometry of both the phenyl ring of azobenzene. We measure the
quality of ML-CG potential by how well they reproduce statistics from the
corresponding CG mapped data (ground truth) by performing CG MD simulations
with a time step of 1 fs at 500 K by placing a single azobenzene molecule in a
cubic box of length 3 nm. For both the AA and noh systems, we have repeated the
same simulation strategy. We compute the RDF (g(r)) of atomic density, as well
as the N=N bond distance, to assess different structural correlations present
in the generated CG trajectory. In the case of azobenzene, the neural network
potential nicely captures the N=N bond length distribution in the CG
simulation. The probability distribution of the N=N bond distance exhibits a
dominant peak at around 0.12 nm in both the CG trajectory and CG-Mapped data,
as shown in Figure \ref{fig:azo}b, depicting a significant overlap between the
distribution of N=N bond length. While in AA and noh systems, the dominant peak
for N=N bond distance appears at 0.13 as shown in Figures 4a and 5a  in the SI.
On the other hand, in the coarsened system, the average distance between C1 and
N bead appears at 0.19 nm in both CG and CG mapped trajectory, as shown in
Figure \ref{fig:azo}c. In case of AA and noh systems, the probability distance
density distribution profile of C1 and N atoms in Figure 4b and 5b in the SI
shows a highly intense peak around 0.14 nm. Furthermore, we conducted a
dihedral angle analysis by defining the dihedral angle between atoms C1-N-N-C1
using the PLUMED plug-in. The 1D projection of the dihedral angle density is
shown in Figure \ref{fig:azo}d. The dihedral density distribution profile shows
excellent agreement between the CG trajectory and CG-Mapped data since in both
cases, the most dominant peak arises in the range of $150^\circ$ to
$180^\circ$. Interestingly, in AA and noh simulations, the ML CG model
excellently recovers the probability distribution of the dihedral angle as
shown in Figures 4c and 5c in the SI. Figure 6c in the SI shows the g(r) of the
atomic density of the coarsened model and CG-Mapped data. There is a nice
agreement between the CG trajectory and the ground truth data, suggesting that
the ML-CG model accurately captures the RDF peak position and intensity of the
ground truth data. Figures 6a and 6b in the SI show close agreement with the
ground truth data in both AA and noh simulations. On the other hand
Jensen-Shannon Divergence (JSD) of RDFs shows close alignment between the CG
trajectory and the CG-Mapped data, as shown in Table IV in the SI. The lower
JSD values in Table IV in the SI suggest the ML-CG model performs better in
each system. These results illustrate the potency of graph neural network
potentials in bottom-up molecular modeling for capturing molecular level
interactions.

\subsection*{3BPA}
\begin{figure}[!ht]
    \centering
    \includegraphics[width=0.6\linewidth]{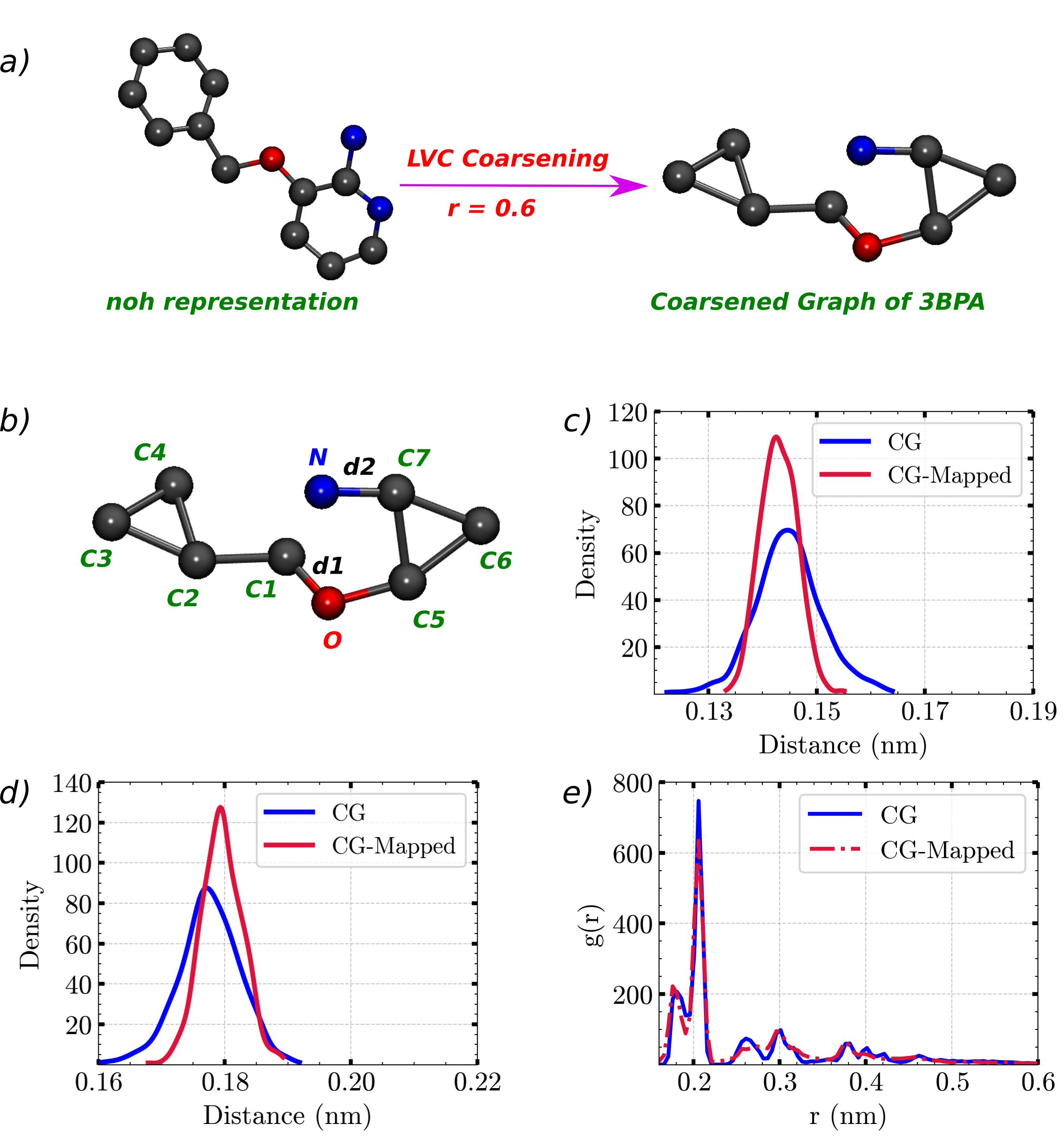}
    \caption{a). Schematic representation of the coarsening strategy for the
3BPA molecule using the LVC coarsening algorithm. Probability distribution of
b). Coarsened atoms and different bond lengths. Probability density
distribution of distance between c). C1 and O atoms d). C7 and N atoms. e).
g(r) of atomic density. Blue line for CG trajectory and red line for CG-Mapped
data} \label{fig:bpa}
\end{figure}
As another example, we selected 3BPA molecule. The dataset contains the
configurations from DFT-MD simulations sampled at 300 K, 600 K, and 1200 K in
gas phase. In this work, we train the MACE architecture on 500 configurations
sampled at 300 K and tested on 1669 configurations sampled at the same
temperature in all cases. First, we have trained our MACE architecture with the
coordinates and forces from AA and noh representations to generate the ML
potential for respective representations. After that, we used the unsupervised
local variational cliques (LVC) algorithm by fixing the coarsening ratio at 0.6
as shown in Figure \ref{fig:bpa}a, b. The noh representations of the 3BPA
molecule and the corresponding forces on each noh atom are used as initial
features for the LVC coarsening algorithm. In the coarsened representation,
each ring is represented by three beads. Furthermore, the N atom of the -NH2
group is represented by a single CG bead. Then, the MACE architecture is
trained with the coordinates and forces present on each node of the CG
representation until the instantaneous force-matching loss is minimized, as
mentioned in the method section. The MACE architecture was trained up to 300
and 100 epochs, respectively, for AA, noh, and CG representation. Further
details about training hyperparameters are provided in Table V SI. Now, we
performed three sets of independent simulations by placing each representation
of the 3BPA molecule in a cubic simulation box of length 5 nm using a timestep
of 1 fs at 300 K. We performed different bond-distance distribution and g(r) of
atomic density analysis to assess different structural correlations in each
system.  We investigate the C-O bond length distribution in the CG trajectory
and compare it with the CG-Mapped trajectory. Interestingly, the C-O bond
distribution (d1) shows good agreement with the CG-Mapped data as shown in
Figure \ref{fig:bpa}c. We also perform C7-N atom distance (d2) analysis as
shown in Figure \ref{fig:bpa}d, demonstrating the capability of the ML-CG
method in reproducing the C7-N bond length distance as in CG-Mapped dataset.
While Figures 7b and 8b show C-O bond length (d1) appears at around 0.14 nm in
both the AA and noh systems and their respective datasets. We further study the
probability distribution of the C12-N bond length in both AA and noh
simulations, as shown in 7c and 8c. In each case, the bond length distributions
are in close alignment with the ground truth data. To emphasize the The RDF
(g(r)) of atomic density of 3BPA molecule in CG simulation nicely reproduced
the g(r) of CG mapped data(ground truth) as shown in  Figure 5e, implying that
the ML-CG potential nicely retain the structural correlations in the CG
resolution as present in CG-Mapped data. On the other hand, Figures 7d and 8d
in the SI show that the g(r) in AA and noh representation nicely capture the
peak position and intensity of g(r) from their respective training dataset. In
addition, the Jensen–Shannon Divergence (JSD) of RDF shows close agreement
between the CG trajectory and the CG-Mapped trajectory (ground-truth data), as
shown in Table VI of the SI.

\section{Applicability and Limitations}
Our study highlights the application of a novel unsupervised multi-level graph
coarsening method with local variation in the lens of the bottom-up CG
perspective. This algorithm omits a key challenge in the graph coarsening, the
definition of a bead mapping protocol which plays a critical in the success of
any CG simulation. Usually, coarse-grained beads were defined by combining a
set of nonhydrogenic heavy atoms into CG sites, with no graph-based grouping ;
the mapping is chosen manually and held fixed during training. By leveraging
the fundamental concepts of graph theory, we can achieve our desired nodes in
the coarsened graph and retain the essential information or the chemical
topology of the original graph. The automated nature of multi-level coarsening
algorithm derives the coarsened graph by merging vertices from the candidate
set based on local variational cost, which is in contrast to the typical human
intuition-based CG-Mapping scheme. Moreover, this multi-level coarsening
algorithm does not involve any learning parameters throughout the coarsening
process. The entire coarsening process is CPU-bound and computationally fast.
Unlike other existing ML coarsening architectures, which require the training
hyperparameters, LVN/LVC require only the information of the coarsening ratio
and edge connectivity. On the other hand, MACE architecture provides a
promising avenue for accurately parameterizing the ML-CG potential without
using the prior potentials by retaining the essential equilibrium properties of
the all-atom representation.


A potential limitation of our current framework is that we do not leverage any
neural network–based graph condensation or learned coarsening schemes. Instead,
we rely solely on the unsupervised Local Variation Neighborhood (LVN) and Local
Variation Clique (LVC) criteria to select contraction sets. While this means we
forego the adaptability of trainable models, it offers distinct benefits: the
mapping requires no training hyperparameters, remains fully deterministic and
interpretable, and incurs minimal computational overhead compared to learned
coarsening approaches.

On cyclic molecular graphs with their characteristic ring motifs, the standard
neighborhood-based LVN can sometimes over-aggregate along linear neighborhoods,
causing subtle distortions of the ring topology. However, this very behavior
reveals an opportunity: by defining contraction candidates as maximal graph
cliques, our method can instead “lock on” to each ring as a single unit,
systematically coarsening cycles in a way that faithfully preserves their
cyclic integrity.

While full ablation study between large molecules is infeasible bacause of
large conformational space in the training dataset. In our future study, we aim
to expand the scope of graph coarsening-based ML-CG for larger proteins and
bio-macromolecules.

\section{Conclusion}
In this study, we demonstrated an ML-CG workflow based on unsupervised
multi-level graph coarsening algorithm by  leveraging the fundamental
principles of graph theory. Using this algorithm, we can derive the coarsen
representations of a large number of organic molecules which include a wide
variety of functional groups and rings by considering the coordinates, forces,
and edge connectivity between heavy atoms. Our approach is totally CPU-bound
and computationally cheaper. For each system this unsupervised multi-level
coarsening method successfully generated CG representation by preserving the
chemical topology of the original graph. In addition, computational speed is a
key aspect in ML-CG force fields. We introduced MACE-CG for constructing ML-CG
potential from coordinates and forces from the CG-Mapped trajectory
(ground-truth dataset). MACE is a popular widely used higher-order equivariant
message-passing architecture in Machine-Learning interatomic potentials
context. By using MACE we directly eliminate the need of prior potential for
ML-CG training purpose. We demonstrate the capability of the proposed ML-CG
model on existing benchmark MD17 dataset. The ML-CG models accurately reproduce
the equilibrium properties of the ground-truth dateset as quantified from
several bond length distributions and RDF. We hope that our proposed ML-CG
workflow will find adoption within the CG chemistry community, particularly for
defining CG beads.

\section*{Acknowledgements}
SM and DB thank IIT Delhi for the Ph.\,D.\ fellowship. SH thanks IIT Delhi for
the PMRF fellowship. SK and TK thank IIT Delhi for the seed grant. We thank the
IITD HPC facility (FIST) for computational resources.

\section*{Author contributions statement}
SM, SH, SK and TK designed the research. SM, SH, DB carried out all simulations
and analyzed the data. SM, SH, SK, and TK wrote the manuscript.

\bibliographystyle{unsrt}
\bibliography{ref}

\end{document}